\documentstyle[emulateapj,psfig]{article}

\received{2000 September 22}
\accepted{}
\journalid{}{}
\articleid{}{}

\slugcomment{To appear in {\it The Astrophysical Journal Letters}}

\lefthead{Mirabal \& Halpern}
\righthead{A Neutron-Star Identification for 3EG J1835+5918}

\begin{document}

%
%
\def\xray{RX~J1836.2+5925}
\def\source{3EG~J1835+5918}
\def\ro{{\it ROSAT\/}}
\def\asca{{\it ASCA\/}}

\title{A Neutron-Star Identification for the High-Energy Gamma-ray\\
Source 3EG J1835+5918 Detected in the {\it ROSAT\/} All-Sky Survey}

\author{N. Mirabal and J. P. Halpern}
\affil{Astronomy Department, Columbia University, 550 West 120th Street,
 New York, NY 10027}
\affil{abulafia@astro.columbia.edu, jules@astro.columbia.edu}
\authoremail{abulafia@astro.columbia.edu, jules@astro.columbia.edu}

\begin{abstract}
\rightskip 0pt \pretolerance=100 \noindent
In the error box of \source,
the brightest as-yet unidentified EGRET source at intermediate
Galactic latitude,
we find a weak, ultrasoft X-ray source at energies $E < 0.3$~keV in the 
\ro\ All-Sky Survey.  Deep optical imaging at the location of this
source, as pinpointed by an observation with the \ro\ HRI,
reveals a blank field to a limit of $V > 25.2$.
The corresponding lower limit on $f_X/f_V$ is 300, which signifies
that the X-ray source is probably a thermally emitting neutron star.
Considering our previous complete multiwavelength survey of the \source\ 
region which failed to find any other notable candidate for identification
with \source, we propose that this X-ray source, \xray, is a rotation-powered
$\gamma$-ray pulsar which is either older or more distant than the
prototype Geminga.  We see marginal evidence for variability between two \ro\
HRI observations.  If real, this would indicate that the X-ray emission has 
an external origin, perhaps due to intermittent heating of the polar
caps by a variable particle accelerator. \xray\ could even be an old,
recycled pulsar, which may nevertheless have a high $\gamma$-ray efficiency.
\end{abstract}


\section{Introduction}

The nature of the persistent high-energy ($> 100$ MeV)
$\gamma$-ray sources in the Galaxy remains an enigma
three decades after their discovery.  Ever since the
identification of the mysterious $\gamma$-ray source Geminga as the
first radio quiet but otherwise ordinary pulsar (see review by
Bignami \& Caraveo 1996), it has been argued
that rotation-powered pulsars likely dominate
the Galactic $\gamma$-ray source population
(e.g., Yadigaroglu \& Romani 1997), and that many of them
will be radio quiet.
That Geminga is a bright EGRET source at a distance of only $\sim 160$~pc
(Carveo et al. 1996) begs for it not to be unique.

Several excellent neutron-star candidates for EGRET sources have
emerged from X-ray observations in recent years.
One is present in the supernova remnant CTA1 which is close to
3EG~J0010+7309 (Brazier et al. 1998).
Another is in the $\gamma$-Cygni supernova remnant which is
coincident with 3EG~J2020+4017 (Brazier et al. 1996).
An intriguing 34~ms X-ray pulsar with a Be star companion 
lies in the error circle of 3EG~J0634+0521 (Kaaret et al. 2000).
X-ray nebulae that are inferred
to contain pulsars have been detected in the error circles of
the EGRET sources 2EG~J1811--2339 (Oka et al. 1999) and
3EG~J1420--6038 (Roberts \& Romani 1998; Roberts et al. 1999).
Radio-quiet neutron stars have been found
which are not apparently EGRET sources, but they are
important as members of the growing class of cooling neutron stars
(RX~J185635--3754, Walter 2000; PKS~1209--52, Zavlin et al. 2000).
Neutron-star candidates discovered by \ro\ were reviewed
by Caraveo, Bignami, \& Tr\"umper (1996) and by Motch (2000).

As the brightest of the as-yet unidentified intermediate-latitude EGRET sources 
at $(\ell, b)=(89^{\circ},25^{\circ})$, \source\ is a good {\it a priori}
target for the next such identification.
We recently completed an exhaustive search for a counterpart
of \source. 
The observations included deep radio, X-ray, and optical surveys,
as well as optical spectroscopic classification 
of every active object within or close to its 99\% confidence
error ellipse (Mirabal et al. 2000, hereafter Paper~1).
In summary, we identified optically all but one of the {\it ROSAT\/}
and {\it ASCA\/} sources in the region of \source\
to a flux limit of
$\sim 5 \times 10^{-14}$~erg~cm$^{-2}$~s$^{-1}$, which is $10^{-4}$
of the $\gamma$-ray flux, without finding any suggestive
evidence for a possible counterpart among the identified
sources.  We also proposed that the
one unidentified X-ray source, \xray, is the most promising
candidate for identification with \source\ principally because
the {\it absence} of an optical counterpart for it is the standard signature
of an isolated neutron star,
or perhaps a more exotic compact object.

The recent release of the decade-old \ro\ All-Sky Survey ({\it RASS\/})
data (Voges et al. 2000) provides an important new tool for detecting
neutron stars by means of their soft, thermal X-ray emission,
and thus for furthering
the search for the counterparts of nearby EGRET sources.
In this Letter, we report new results
from the {\it RASS} on the previous candidate \xray\ which 
greatly enhance its
credentials as the likely pulsar counterpart of \source.  We also present
a detailed study of the X-ray and optical objects in its immediate
neighborhood which establishes that \xray\ is indeed undetected to a limit of
$V > 25.2$, as would be expected for a neutron-star counterpart of \source.

\section{X-ray Observations of \xray\ }

A total of four X-ray observations that cover the entire
99\% error ellipse of \source\ have been made 

\bigskip
\centerline{
\psfig{file=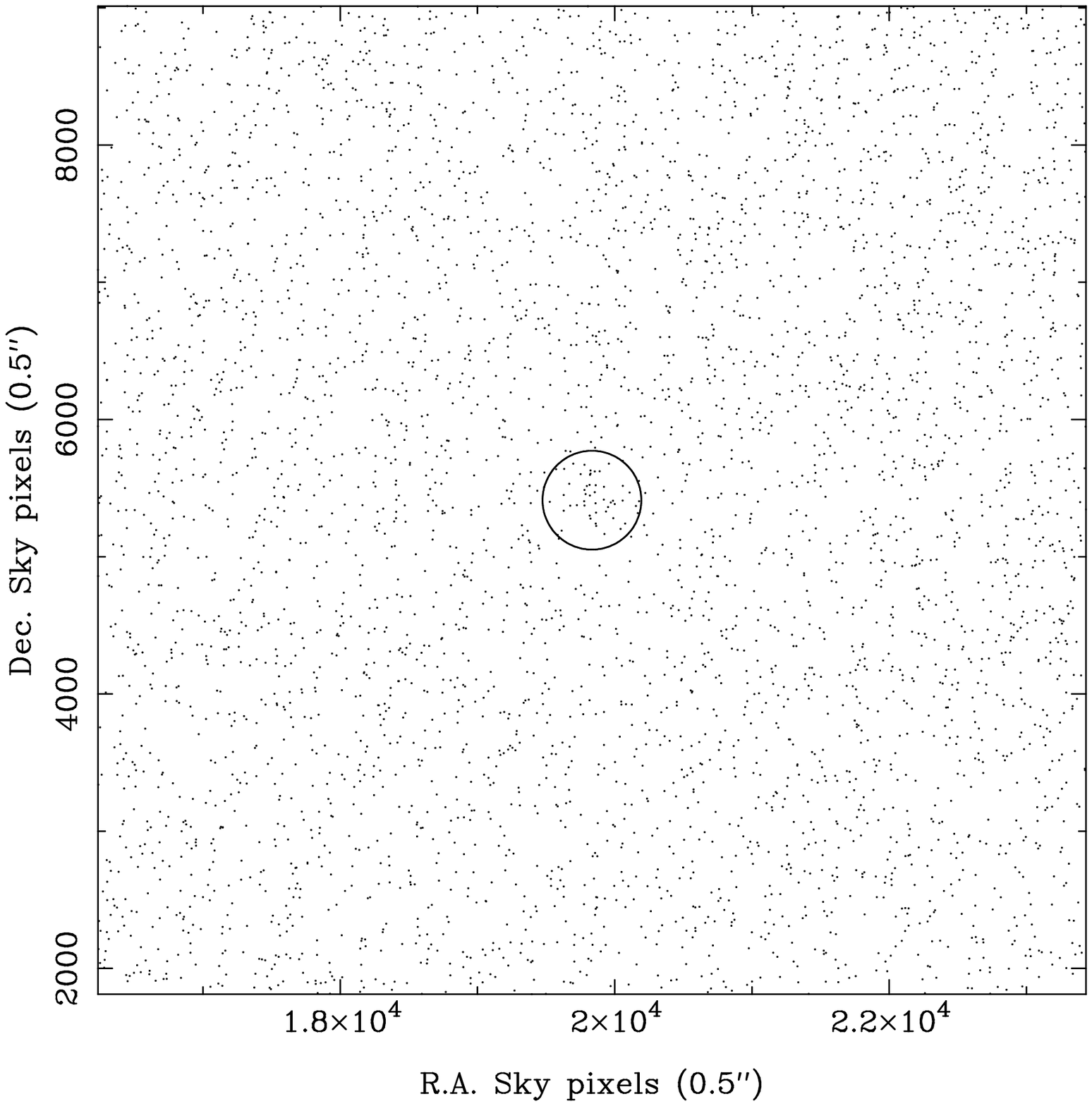,height=3.0in}
}
\medskip
{\footnotesize  FIG. 1.--- A photon map of the $1^{\circ}\times 1^{\circ}$ region
around \xray\ from the {\it ROSAT\/} All-Sky Survey.  Only events
with pulse heights in the range 0.1--0.35~keV are shown.  The net,
background subtracted counts from \xray\ number $22.3 \pm 6.3$.
}
\bigskip

\noindent
to date, one by the {\it RASS\/} with the PSPC instrument,
two by the \ro\ High Resolution Imager (HRI),
and one by the \asca\ GIS.  The dates and exposure times of
these observations are listed in Table~1.  The results
of the two pointed \ro\ observations and the one \asca\
observation were presented in Paper~1, and
they will not be repeated here except where relevant to
the properties of \xray.

In the {\it RASS\/} Faint Source Catalog, we find an entry
at the position (J2000) $18^{\rm h}36^{\rm m}13^{\rm s}\!.59,\
+59^{\circ}25^{\prime}30^{\prime\prime}\!.5$ with an
uncertainty of $26^{\prime\prime}$ and a count rate
of $0.0146 \pm 0.0041$~s$^{-1}$.  The effective exposure time at this location
was 1532~s.   The hardness ratio of the source is given as $-1.0$, meaning
that all of its photons fall at energies
below 0.4~keV.  This source is consistent in position and flux 
with the later observed \ro\ HRI source \xray\ as described in Paper~1,
and since it is by far the brightest source within
a region of radius $10^{\prime}$ there is no reason to suspect confusion
with the many fainter HRI sources in its vicinity.  An image of the
source and its surrounding field in the {\it RASS\/} is shown in Figure~1.

A deep X-ray image of the field around \xray, taken from the longer \ro\
HRI observation, is shown in Figure~2.  \xray\ is the brightest
source in this Figure.  Six of the fainter X-ray sources
were identified
using multicolor CCD imaging and optical spectroscopy.
Their positions and counterparts are given in Table~1 of Paper~1.
We use these sources to tie the X-ray positions to the optical astrometric
reference frame employed by the USNO--A2.0 catalog (Monet et al. 1996).
Figure~3 shows the offsets between the X-ray positions and the optical
positions before and after a zero-point shift was applied to the X-ray 
positions in order to correct a 
small systematic error which is hardly
apparent in the individual identifications.  This offset is only
$1^{\prime\prime}\!.3$ in right 
ascension and $0^{\prime\prime}\!.8$ 
in declination.  The final error circle

\bigskip
\centerline{
\psfig{file=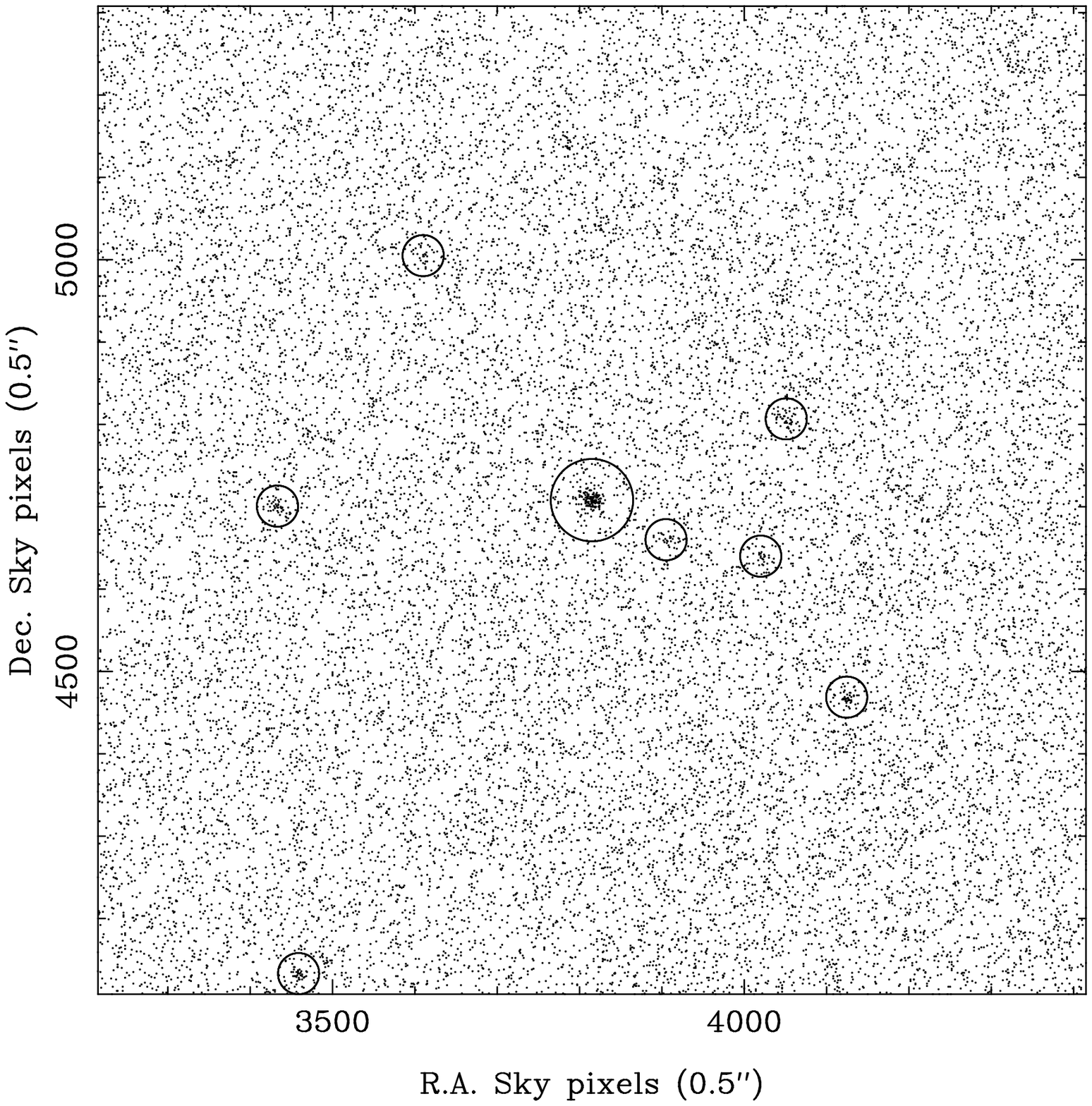,height=3.0in}
}
\medskip
{\footnotesize FIG. 2.--- A $10^{\prime}\times 10^{\prime}$ portion of the \ro\ HRI image
centered on the brightest, unidentified X-ray source \xray\
({\it large circle}).
As refined with the help of Figure~3, the position of \xray\ is
(J2000) $18^{\rm h}36^{\rm m}13^{\rm s}\!.77,\
+59^{\circ}25^{\prime}30^{\prime\prime}\!.4$.
}
\bigskip

\noindent
radius of $3^{\prime\prime}$
includes all of the X-ray source identifications in Figure~3, and it is
undoubtedly a conservative estimate of the positional error of the bright
unidentified source \xray.  The corrected position for the latter is
(J2000) $18^{\rm h}36^{\rm m}13^{\rm s}\!.77,\
+59^{\circ}25^{\prime}30^{\prime\prime}\!.4$.  We adopt this as the
best X-ray position.  It differs by only
$1^{\prime\prime}\!.5$ from the position given in Paper~1,
and it agrees very well with the position of the corresponding {\it RASS}
source.

We further examined the nature of \xray\ in the {\it RASS\/}
by extracting source photons from within a radius of $3^{\prime}$,
and obtaining a background estimate from an annulus of inner radius
$4^{\prime}$ and outer radius $8^{\prime}$.  Although this source contains
only 22 net photons it is clearly real; Voges et al. (2000)
give a probability of $3 \times 10^{-7}$ that it is spurious.
By binning the photons
into pulse-height intervals of 0.1~keV, we see that there is no evidence
for any emission above 0.3~keV (see Table~2).  If fitted by a blackbody model,
such a pulse-height distribution is consistent with
$T \leq 5 \times 10^5$~K, but it is also dependent
upon the unknown intervening column density.  If we assume
$1 \times 10^{20} < N_{\rm H} < 3 \times 10^{20}$~cm$^{-2}$, the bolometric
flux corresponding to an assumed $T = 5 \times 10^5$~K is in the range
$(1.5-5.7) \times 10^{-13}$ ergs~cm$^{-2}$~s$^{-1}$.  The soft spectrum
of this source also explains why it was not detected in the \asca\ observation,
which had a 1--10~keV detection limit of
$\sim 1 \times 10^{-13}$ ergs~cm$^{-2}$~s$^{-1}$.

The net count rate of \xray\ in the \ro\ HRI observation of 1998 is
$(2.25 \pm 0.23) \times 10^{-3}$~s$^{-1}$.  For the spectral parameters
assumed above, the HRI flux agrees with the PSPC flux measured 7.5~yr
earlier to better than 10\%.  Although this is consistent with
a thermal neutron star interpretation, the source was possibly {\it not}
detected in the shorter HRI observation in 1995.  The formal
count rate in the latter observation was only
$(0.90 \pm 0.51) \times 10^{-3}$~s$^{-1}$.  Because of the small
number of photons detected, the significance of the variability
should be evaluated using Poisson statistics.  In this analysis we 

\centerline{
\psfig{file=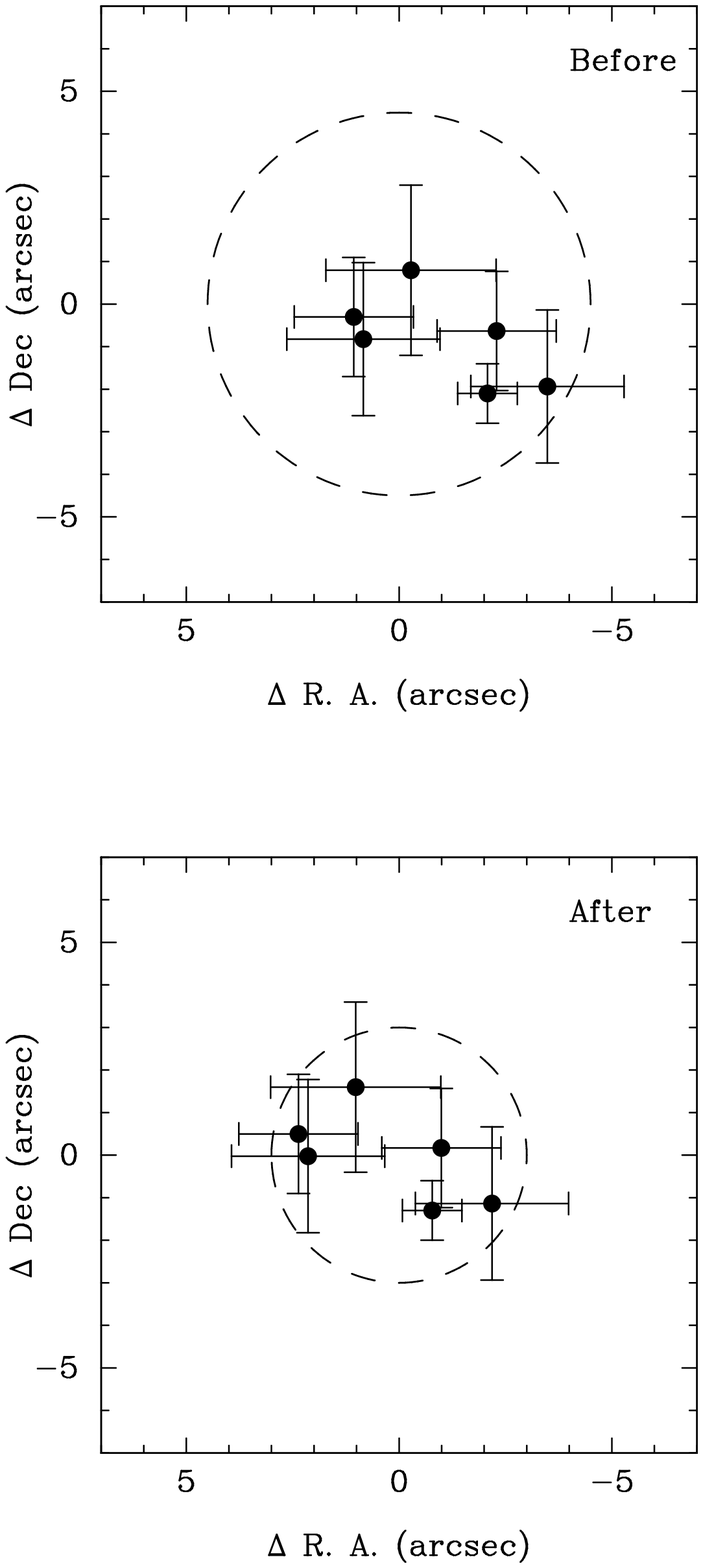,height=6.0in}
}
\medskip
{\footnotesize FIG. 3.--- The result of using six optically identified, fainter HRI sources
from Figure~2 to tie the X-ray positions to the optical astrometric 
reference frame.  {\it Top}: The uncorrected position offsets.
{\it Bottom}: The final offset between X-ray and optical position after
applying a zero-point shift of $1^{\prime\prime}\!.3$ in right ascension
and $0^{\prime\prime}\!.8$ in declination.
Error bars are the 90\% X-ray statistical uncertainties.
We adopt a conservative radius of $3^{\prime\prime}$ for
the uncertainty in position of \xray.
}
\bigskip

\noindent
use pulse-height channels 1--9, as there are virtually no
X-ray photons in higher channels in the HRI. If we assume a constant
source given by the count rate in the longer HRI observation, then
we should expect to find in the source detection circle in 1995 a total
of 24 photons (20 source and 4 background), whereas
a total of only 12 were detected.
The Poisson probability of obtaining 12 or fewer events when 24 are expected
is $5.4 \times 10^{-3}$.  Thus, the source is variable at the 99.5\%
confidence level.  However,
the existence of a \ro\ PSPC flux which is consistent with
the latest HRI detection may be regarded as evidence contradictory to
this indication of variability,
and we would not be surprised if the one weak HRI
detection turns out to be a statistical anomaly.
Furthermore, the source {\it was}
persistent during the 1 month span of the HRI observation in which it
was detected.  The existence or not of variability
is crucial to the detailed interpretation of the physics of
this source, as we discuss below,
but not to its identification with \source.

\section{Optical Observations of \xray\ }

Armed with the precise X-ray position of \xray,
we reexamined the deepest optical images which we obtained
on the MDM Observatory 2.4m telescope.
Figure~4 is a reproduction of the $V$-band image from
Paper~1, a total of 2~hr of exposure obtained on 2000 July 24.
The adopted $3^{\prime\prime}$ radius error circle superposed.
The circle is blank to a $3\sigma$ limit of $V > 25.2$.  A limit of
$R > 24.5$ was also obtained on 2000 July 15 (see Paper 1).

In order to guard against the possibility of an anomalous
systematic error in the X-ray position, we obtained
optical spectra of the three nearest objects to the west of the error circle
on the Palomar 5m telescope.  These objects are a late K star of
magnitude $V = 20.7$, and two faint galaxies which show no evidence of
activity in their optical spectra.  They are therefore
not viable candidates
for identification with \xray.
We have also determined that no optical object
in this field shows proper motion which could account
for its positional discrepancy with the X-ray source.
Thus, \xray\ remains undetected
optically to a limit of $V > 25.2$, even allowing for
a conservative uncertainty 
on its position.  This upper limit implies that
the ratio of X-ray-to-optical flux $f_X/f_V$ is greater than 300,
an extreme which is seen only among neutron stars.  

\centerline{
\psfig{file=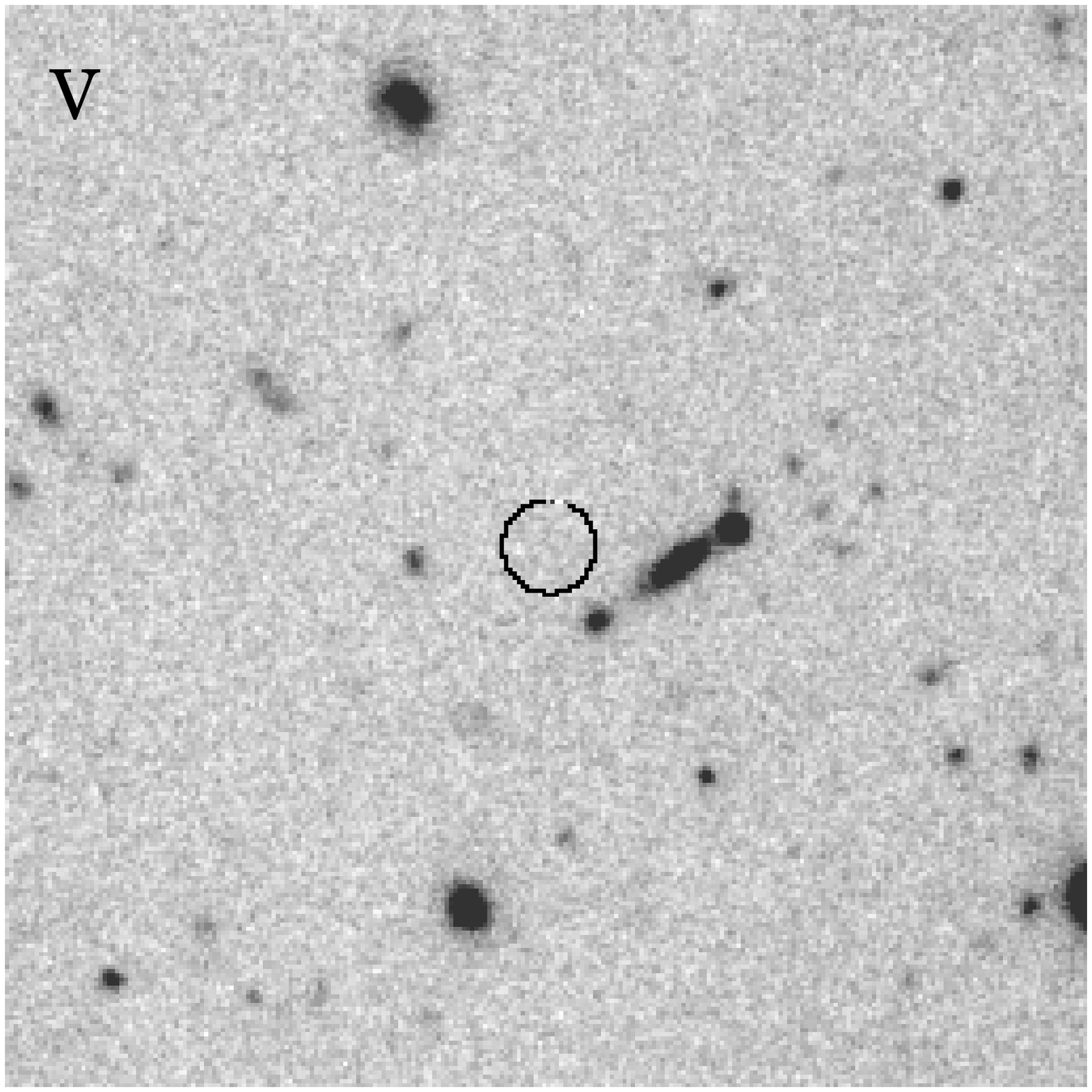,height=3.8in}
}
{\footnotesize FIG. 4.--- A deep $V$-band image at the location of the unidentified
X-ray source \xray.  This 2~hr summed exposure from
the MDM 2.4m telescope in seeing of $0^{\prime\prime}\!.8$ yielded
a $3\sigma$ detection limit of $V = 25.2$.  North is up, and east
is to the left. The field is $70^{\prime\prime}$ across, and
the \ro\ HRI error circle is drawn as a conservative
$3^{\prime\prime}$ in radius as derived from Figure~3.
The best position is (J2000) $18^{\rm h}36^{\rm m}13^{\rm s}\!.77,\
+59^{\circ}25^{\prime}30^{\prime\prime}\!.4$.
}

\section{Interpretation and Conclusions}

The latest analysis of the EGRET
observations of \source\ leads to the conclusion that it shows no
evidence for long-term variability (Reimer et al. 2000).  Its spectrum can
be fitted by a relatively flat power law of photon index
--1.7 from 70~MeV to 4~GeV, with a turndown above 4~GeV.
Such temporal and spectral behavior are more consistent with a
rotation-powered pulsar than a blazar, which is the other major
class of EGRET source.  
The soft X-ray source \xray\ located 
in the error box of the \source\ 
presents properties that resemble a neutron star, and
could be the second case of a nearby, radio-quiet 
$\gamma$-ray pulsar.

The detection of \xray\ in the {\it RASS} 7.5~yr prior to the
final HRI observation eliminates an alternative hypothesis mentioned
in Paper~1, that it might have been a luminous soft X-ray transient
caused by the tidal disruption of a star by a supermassive black hole
in a distant galaxy, and unrelated to \source.
Such events are expected (and observed) to last
only a few months (e.g., Komossa \& Bade 1999; Komossa \& Greiner 1999),
while this source has persisted for at least 7.5~yr.
Furthermore, our deep optical imaging failed to find a candidate galaxy
at this location.

The available data from radio through $\gamma$-rays 
are thus consistent with the hypothesis that \xray\ is the actual
counterpart of \source\ and a more distant or older cousin of the
Geminga pulsar.   
The implications of a neutron-star interpretation of \xray\
were discussed at length in Paper~1.  Although the intrinsic
flux of the source is highly uncertain because of the
unknown column density, the latter is probably not greater than
$3 \times 10^{20}$~cm$^{-2}$ or we would not be seeing such a soft
spectrum.  Then \xray\ is evidently at least 10--40 times
fainter than Geminga.  Thus it is either more distant than Geminga
($d > 160$~pc, Caraveo et al. 1996),
or cooler ($T < 5 \times 10^5$~K, Halpern \& Wang 1997).
But a cooling neutron star of age $< 10^6$~yr should not be
{\it too} distant, because to place it $> 400$~pc above the Galactic
plane at $b = 25^{\circ}$
would require a kick velocity at birth of $> 500$~km~s$^{-1}$.
A reasonable upper limit on the distance is therefore 1~kpc,
which we note implies an isotropic $\gamma$-ray luminosity of
$6 \times 10^{34}\,(d/1\,{\rm kpc})^2$ ergs~s$^{-1}$, rather
more than the spin-down power $I\Omega\dot \Omega$ of Geminga
($3.3 \times 10^{34}$ ergs~s$^{-1}$).  If the actual spin-down power
of \source\ is determined, it will place an additional constraint
on the distance and beaming factor.

The only observation that possibly complicates this interpretation is the
marginally significant long-term (and non-monotonic) variability
by about a factor of two.  Such behavior is not consistent with
surface thermal radiation of the heat of formation from the interior.
However, it is possible that thermal emission from older neutron stars
is dominated by the heated polar caps impacted by inflowing relativistic 
particles from the acceleration regions which are responsible for their
pulsar action.  Indeed, such reheating
has been shown to be the source of the X-rays from
the nearby recycled millisecond pulsar PSR~J0437--4715
(Zavlin \& Pavlov 1998).
If under such conditions the particle acceleration regions in the pulsar
were intermittent, the thermal X-rays resulting from this external heating 
would also be variable.  
In principle, \source\ could be as powerful as Geminga at $\gamma$-ray
energies, but less conspicuous at soft X-ray wavelengths because its
surface has cooled except for the $\sim 1$~km radius polar caps which
are being reheated directly by the accelerator.
The most efficient $\gamma$-ray pulsars are expected to be the ones
like Geminga, operating closest to their death lines (Chen \& Ruderman 1993).
A millisecond pulsar typically possesses spin-down power and
magnetospheric gap voltage similar to Geminga and other ordinary
pulsars.  Since there is also evidence that the magnetospheric
X-rays from Geminga are variable (Halpern \& Wang 1997),
either an exotic middle-aged pulsar 
or a maximally efficient recycled pulsar are plausible scenarios for \source.

\acknowledgments{We thank Peter Mao and Fiona Harrison for obtaining
the optical spectra mentioned in \S 3.}  This work was supported by
NASA grant NAG 5--9095.  

\begin{deluxetable}{llcr}
\tablenum{1}
\tablecolumns{4}
\tablewidth{0pc}
\tablecaption{X-ray Observations of \xray/\source }
\tablehead
{
            & \omit \hfil Dates \hfil & Exposure Time  & \omit \hfil Count Rate \hfil  \\
Instrument  & \omit \hfil (UT) \hfil & (s) & \omit \hfil (s$^{-1}$) \hfil \\
}
\startdata
\ro\ PSPC\tablenotemark{a} &  1990 Oct 11--24		 &  1532  & $(1.46 \pm 0.41) \times 10^{-2}$ \\
\ro\ HRI    		&  1995 Feb 2--4		 &  9078  & $(0.90 \pm 0.51) \times 10^{-3}$ \\
\ro\ HRI   		&  1997 Dec 15 -- 1998 Jan 20	 & 60603  & $(2.25 \pm 0.23) \times 10^{-3}$ \\
\asca\ GIS  		&  1998 April 20--22		 & 68900  & \omit \hfil . . . \hfil \\
\enddata
\tablenotetext{a}{ The \ro\ All-Sky Survey }
\end{deluxetable}

\begin{deluxetable}{cr}
\tablenum{2}
\tablecolumns{2}
\tablewidth{0pc}
\tablecaption{{\it RASS\/} Spectrum of \xray\ }
\tablehead
{
Energy bin              & Counts \\
\omit \hfil (keV) \hfil &  \\
}\startdata
0.1--0.2 & $10.4 \pm 5.8$ \\
0.2--0.3 & $11.9 \pm 6.4$ \\
0.3--0.4 & $-0.5 \pm 4.0$ \\
0.4--0.5 & $ +1.0 \pm 3.3$ \\
0.5--2.0 & $-0.6 \pm 5.3$ \\
\enddata
\end{deluxetable}

\end{document}